# Measurement of the background in Auger-photoemission coincidence spectra (APECS) associated with inelastic or multi-electron valence band photoemission processes

S. Satyal[1], P.V. Joglekar[1], K. Shastry[1], S. Kalaskar[1], Q. Dong[2], S.L. Hulbert[2], R. A. Bartynski[3], and A.H. Weiss[1]

[1]The Department of Physics, University of Texas at Arlington, Arlington, TX, 76019. USA
[2]Photon Sciences Directorate, Brookhaven National Laboratory, Upton, NY, 11973. USA
[3]Department of Physics and Astronomy, Rutgers University, Piscataway, NJ 08854, USA

**ABSTRACT**

Auger Photoelectron Coincidence Spectroscopy (APECS), in which the Auger spectra is measured in coincidence with the core level photoelectron, is capable of pulling difficult to observe low energy Auger peaks out of a large background due mostly to inelastically scattered valence band (VB) photoelectrons. However the APECS method alone cannot eliminate the background due to valence band photoemission processes in which the initial photon energy is shared by two or more electrons and one of the electrons is in the energy range of the core level photoemission peak. Here we describe an experimental method to determine the contributions from these background processes and apply this method in the case of Copper $M_3VV$ Auger spectrum obtained in coincidence with the $3p_{3/2}$ photoemission peak. A beam of 200 eV photons was incident on a Cu(100) sample and a series of coincidence measurements were performed using a spectrometer equipped with two cylindrical mirror analyzers (CMAs). One CMA was set at series of fixed energies that ranged between the energy of the core and the VB peaks. The other CMA was scanned over a range corresponding to electrons leaving the surface between 0eV and 70eV. The set of measured spectra were then fit to a parameterized function which was extrapolated to determine the background in the APECS spectra due to multi-electron and inelastic VB photoemission processes. The extrapolated background was subtracted from the APECS spectrum to obtain the spectrum of electrons emitted solely as the result of the Auger process. A comparison of the coincidence spectrum with the same spectrum with background removed shows that in the case of Cu $M_3VV$ the background due to the inelastic scattering of VB electrons is negligible in the region of the Auger peak but is more than half the total signal down in the low energy tail of the Auger peak.

## I. INTRODUCTION

Auger spectra obtained using electron or X-ray induced Auger electron spectroscopy (EAES and XAES) are associated with a large background which is particularly problematic for lower energy Auger peaks (< ~100eV). In the case of X-ray excited Auger, this background is primarily due to photo-emitted valence band (VB) electrons that share energy with other VB electrons resulting in a cascade of secondary electrons. Other contribution includes the background arising from a similar cascade associated with photo-emitted core electrons. Experiments have shown that Positron Annihilation Induced Auger Spectroscopy (PAES) can eliminate secondary electrons due to the impact of the incident beam by using a beam of positrons whose energy is below the secondary electron threshold [1]. In PAES the Auger transitions are excited through the annihilation of core electrons with positrons trapped in a surface localized state providing PAES with a high degree of surface specificity [2]. In the case of photon induced Auger excitation, Auger photoelectron coincidence spectroscopy, (APECS), originally developed by Haak and Sawatzky in 1978 [3] and adapted for a synchrotron light source by Jensen et al. [4], is capable of pulling difficult to observe low energy Auger peaks out of the large background of VB photoelectrons by measuring the Auger spectra in coincidence with electrons emitted in the energy range of the peak of the photo-emitted core electron. In addition, the coincidence requirement results in significantly higher surface selectivity than conventional Auger methods and allows for the experimental separation of complex spectra into component spectra associated with Auger transitions involving specified core hole initial states [5-10, 12, 13].

While the coincidence requirement in APECS excludes most of the background due to VB and core level photoelectrons that share energy with other VB electrons, it is not sufficient to eliminate events in which the incident photon energy is shared among several VB electrons one of which is in the same energy range as the selected core-level photoemission peak. A consideration of the contributions from VB photoelectrons emitted (as a result of inelastic scattering or multi-electron photoemission) in the energy range of the core level photoemission peak suggests that contributions from this process, termed here *VB-VB Coincidence,* could constitute a significant background in the low energy part of APECS spectra. In the case of 200eV photons incident on Cu(100) the tail of the VB photoemission spectrum constitutes about half of the spectra weight in the energy range of the core level peak.

In this paper, we present an experimental method to determine the line shape and magnitude of the background due to VB-VB Coincidence processes (as defined above) to the APECS spectrum and apply it in the case of the $M_3VV$ Cu(100) Auger spectrum taken in coincidence with the $3p_{3/2}$ core photoelectrons. We note that our measurements of the photon induced VB-VB coincidence processes are closely related to electron induced secondary electron measurements reported by Werner et al. [14] in which they have obtained the spectrum of secondary electrons in coincidence with electrons that were backscattered from the surface with a fixed energy loss relative to the incident electron beam energy.

Our work was motivated by previous measurements of the low energy tail of CVV Auger peaks ,(including APECS measurement of the Al LVV peak [10] and PAES measurements of the Cu MVV and Au NVV Auger transitions [11]), which showed significant spectral weight in the low energy tail (LET) of the Auger peaks. Jensen et al. suggested that the large intensity in the LET region in the APECS measurements (~2/3 the intensity of the peak) could be an indication that a significant number of the core-holes were relaxing through multi-electron Auger decays [15]. Jensen et al measured the APECS spectra over a range of ~30eV including the Auger peak and ~15eV below the peak. Here we present measurements of the APECS spectrum of the Cu $M_3VV$ Auger peak over a range of 70eV including the Auger peak and the associated LET ~50eV below the peak (all the way down to 0eV).

Our analysis of the photoemission spectra indicated that the background due to VB-VB coincidence processes was likely to be significant (as much as half the APECS signal in our case) and that consequently it was important to develop a method to determine both the magnitude and line shape associated with this background. To ascertain the line shape and magnitude of the VB-VB contributions to the APECS spectra we made a series of measurements of the electron spectra taken in coincidence with electrons detected at energies above the core-level photoemission peak and below the VB photoemission peak. The coincidence spectra thus obtained contained contributions only from VB-VB processes and not from Auger processes. We then used an extrapolation procedure to determine the VB-VB contributions to the spectra of electrons taken in coincidence with electrons emitted in the energy range of the $3p_{3/2}$ core level. We subtracted our experimentally determined VB-VB spectral contributions from our (APECS) measurements made in coincidence with the (Cu $3p_{3/2}$) core level to obtain the first measurements of the spectra of electrons emitted from the surface due solely to a selected (Cu $M_3VV$) photon induced Auger transition over the full range of emitted energies from the Auger peak down to 0eV.

## II. EXPERIMENTAL SETUP

The experiments were performed at beam line U1A [9] of the National Synchrotron Light Source (NSLS) at Brookhaven National Laboratory using a coincidence spectrometer system designed to allow APECS measurements to be performed at the NSLS. The spectrometer (described in detail elsewhere [10]), consists of two double-pass cylindrical mirror energy analyzers (CMAs) aligned to detect electrons from the same region of a sample illuminated by photons from the monochromator. Coincidence electronics are used to select events in which both electrons are due to the adsorption of the same photon. During data taking, one analyzer (referred to here as the "fixed" analyzer) was set at a fixed energy while the other analyzer (referred to as the "scanned" analyzer) was scanned over a range of energies as appropriate for the desired measurement.

The measurements were performed on a Cu(100) single crystal sample that was cleaned periodically by Argon bombardment sputtering for 25 minutes followed by annealing until the temperature reached ~900° K. Cleaning cycles were repeated until no significant contamination was detected by Auger survey scans. A count rate dependent correction (less than 10% at the count rates used in the measurements) was made to the coincidence spectra using methods explained in detail in [16]. The fixed and scanned analyzers were set at constant pass energies of 300eV and 80eV, respectively, resulting in an energy resolution of 4.8eV and 1.3eV, respectively. All coincidence spectra were taken at room temperature with photon energy of 200eV and the sample biased at -15V DC. The bias voltage increases the absolute kinetic energies by 15eV in order to minimize the impact of stray magnetic fields on the low energy portion of the photoelectron spectra. A conventional, non-coincidence photoemission spectrum was acquired simultaneously along with an accidental background that was subtracted using the techniques outlined in [10].

## III. RESULTS AND DISCUSSION

The wide-range photoelectron spectrum from Cu(100) sample shown in Fig. 1-d contains the following principal features: the valence band, 3p and 3s core levels, MVV auger peaks as well as the secondary electron tail which peaks at the low kinetic energy region. The spectrum was obtained by irradiating a 200eV photon beam on the sample while it was biased with -15V. To measure the contribution due to these VB-VB Coincidences, a series of spectra were obtained in coincidence with electrons at 5 different energies (indicated by arrows in Fig. 1 and tabulated in Table 1.) in the region above the $3p_{3/2}$ core level and below the VB photoemission peak. In this region, only the VB-VB coincidence electrons contribute. The VB-VB Coincidence spectra for the five choices of fixed energies are shown in Fig. 2, panel a-e. The APECS spectrum associated with the $M_3VV$ Auger peak was obtained under the same conditions by setting the energy of the fixed analyzer at the $3p_{3/2}$ photoelectron peak (136.25eV) and is shown in Fig. 2f. This spectrum contains contributions from electrons from both the $M_3VV$ Auger transition emitted in coincidence with $3p_{3/2}$ core photoelectrons as well as electrons emitted via VB-VB processes.

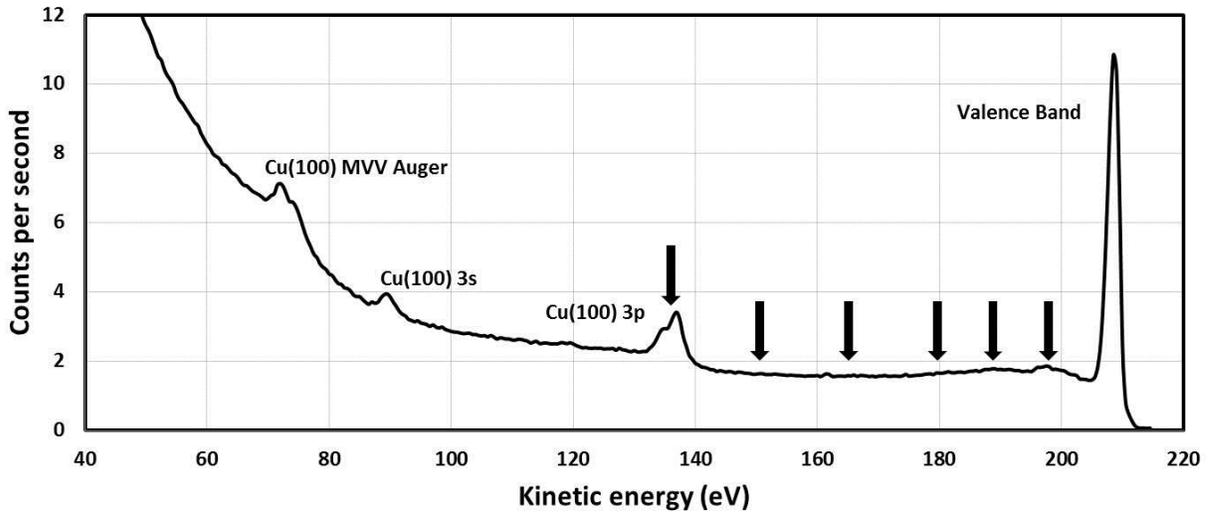

Fig. 1 *Conventional (non-coincidence) X-ray photoelectron spectrum from the Cu(100) sample surface showing the $3p_{1/2}$ and $3p_{3/2}$ core level peaks and the $M_{2,3}VV$ Auger peaks. The spectrum was taken using 200eV photon beam with the sample biased at -15 V DC. The fixed analyzer was set at the energies shown by the vertical arrows while measuring the contributions due to the VB-VB coincidence.*

It may be seen from the series of spectra shown in Fig. 2 panels *a* through *e* that both the width and integrated intensity of the coincidence spectra increase as, $\Delta E$ defined here as the energy difference between the center energy of the fixed analyzer and the energy of the VB photoemission peak, is increased. The panels also show a cut off at an energy $KE_{2max}$ calculated as described below and tabulated in Table 1.

The measured spectra (Fig. 2, a-e) are consistent with the interpretation of the VB-VB mechanism given in the discussion that follows. Here we refer to the kinetic energy of electron detected by the "fixed" analyzer as $KE_1$ and the energy of the electron detected by the scanned analyzer as $KE_2$. If an electron detected in the fixed analyzer at energy $KE_1$ is below the energy of the VB photoemission peak, then it must have shared an amount of energy, $\Delta E$, given by equation 1.

$$\Delta E = h\nu - E_{v1} - \phi_A - KE_1, \qquad (1)$$

where $KE_1$ represents the energy of the detected photo-emitted valence band electron, $h\nu$ is the energy of the incident photon, $E_{v1}$ is the binding energy of the outgoing electron referenced to the top of the Fermi level, $\phi_A$ is the work function of the analyzer. If we assume that the energy, $\Delta E$, is shared with 1 or more additional VB electrons then energy conservation considerations result in an upper limit given by eq. 2 for the energy $KE_2$ of another VB electron detected in coincidence.

$$KE_2 \leq \Delta E - E_{v2} - \phi_A, \qquad (2)$$

where $E_{v2}$ is the binding energy of the detected electron. The equality holds for the case of a two-electron process in which all the energy of the initial photon is shared between the two electrons (either as a result of some direct multi-electron process or through an inelastic collision of the second electron as it exits the surface). The less than relation would apply to multi-electron or energy loss processes in which the energy is shared by more than one additional electron.

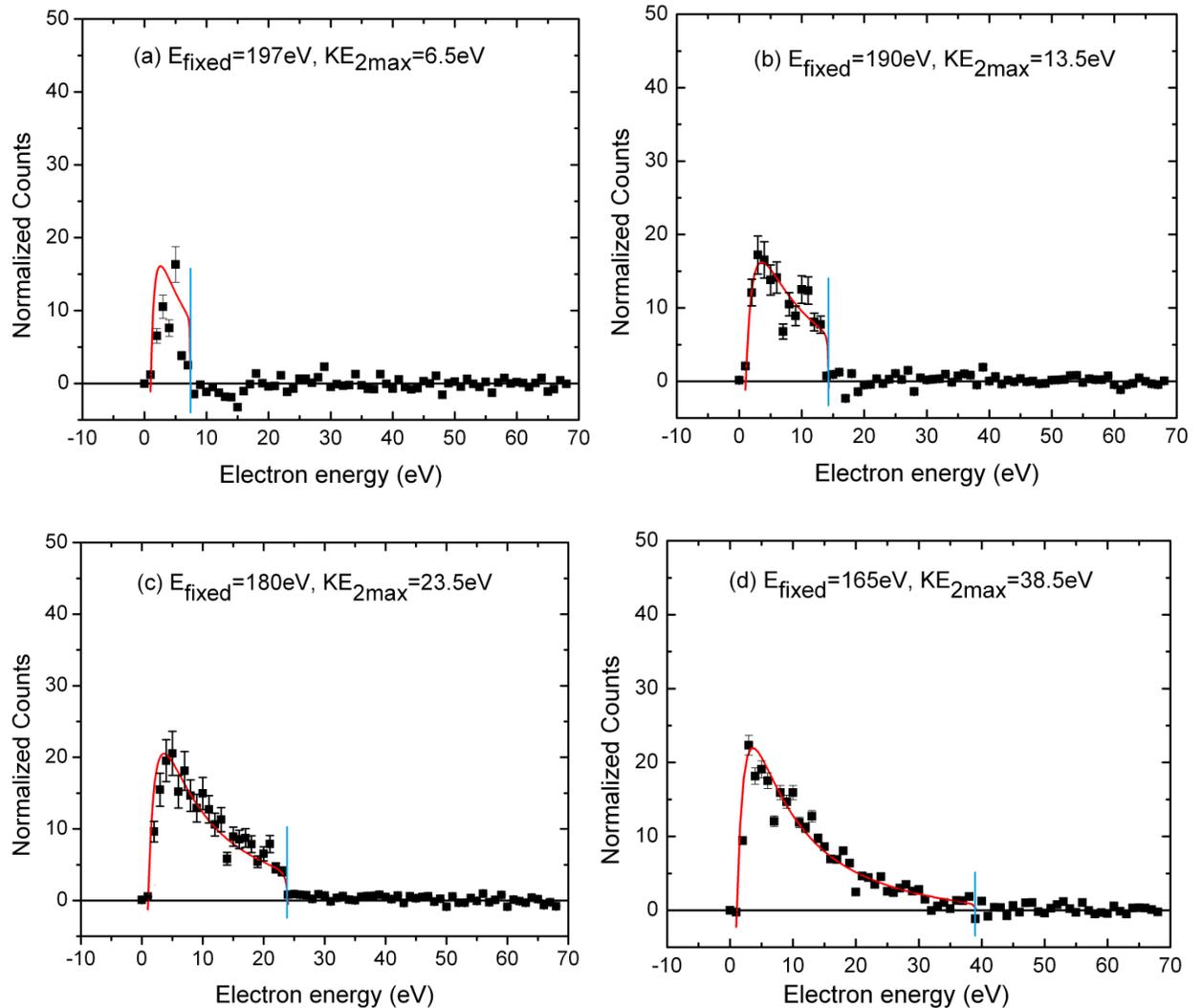

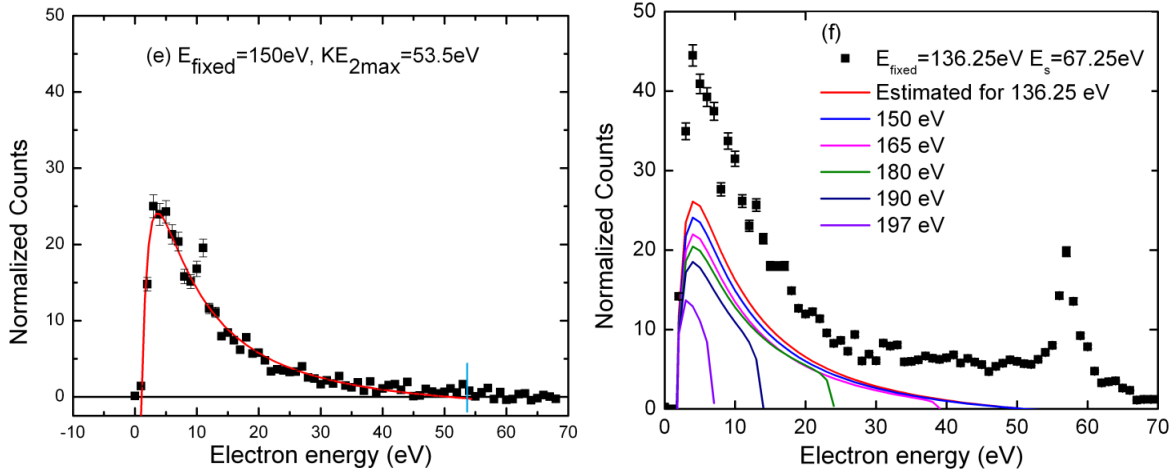

*Fig. 2 Energy spectra taken in coincidence with inelastically scattered VB photoelectrons measured at fixed energies (150eV to 197eV) for Cu(100). The y-axis represents the coincidence counts divided by a number proportional to the total photon fluence and the spectra have been shifted by 15eV to account for the sample bias (Figs. 2-4). The solid red line (panel a-e) represents a fit to a parameterized function discussed in detail below. The blue vertical lines represent the estimated upper energy cutoff of each spectrum at values given in Table 1. The top red curve in panel (f) shows the estimation from the VB-VB contribution for the $M_3VV$ APECS spectrum.*

Table 1 lists the energies of the fixed analyzer used in the coincidence measurements shown in Fig. 2, and the estimated cut off energy $KE_{2max}$. The value of $KE_{2max}$ was calculated using an expression given by eq. 3

$$KE_{2max} = KE_{VB-Peak} - KE_1 - \phi_A, \qquad (3)$$

that can be justified by substituting eq. 1 into eq. 2 and noting that the energy of the VB peak is given by:

$$KE_{VB-Peak} = h\nu - \overline{E_{v1}} - \phi_A, \qquad (4)$$

where $\overline{E_{v1}}$ is close to the peak in the VB density of states.

To facilitate the extrapolation, and guided by the similarities in physics to secondary electron emission, we used a parameterized fitting function previously formulated by Ramaker for use in describing the energy spectra of secondary and redistributed primary electrons emitted as a result of an incident electron beam [10, 18, 19],

$$B(E) = \frac{AE}{(E+D)(E+\Phi)^m} + \frac{B\ln[(E_p-E)/E_b]}{[(E_p-E)/E_b]^n} + C, \qquad (5)$$

where $B(E)$ is the parameterized fitting function and $A$, $B$, $C$, and $D$ are parameters obtained from a nonlinear least square fit of $B(E)$. The constants $m$, $n$ and $E_b$ were taken to be fixed at 1.6, 2, and 4.5eV respectively. The constant $E_p$, corresponding to the energy of the primary electron beam, was fixed at $E_p = KE_{2max}$ as tabulated in Table 1. Our choice of the value of $E_p$, was motivated by making the assumption that the energy available in the VB-VB coincidence process, $E_{available} = \Delta E \approx KE_{2max} + \Phi$ can be treated as a source of available energy comparable to the energy shared by an external primary electron of energy $E_p$ (relative to the vacuum) as it enters the surface, loses energy and drops into a state at the Fermi-level ($E_{available} = E_p + \Phi$). In Fig. 2, the red line represents the background fit function modeled from Eq. 5. The parameters $A$, $B$, $C$, and $D$ are derived from the fits to the individual inelastic VB spectra. We used quadratic fits to the values of these parameters as a function of $\Delta E$, to find extrapolated parameters which were used to obtain an extrapolation of the inelastic VB spectra at $\Delta E = 67.25$ eV corresponding to setting the fixed analyzer at the 4p core photoemission peak. Table 1 shows the shared energy ($\Delta E$) values calculated for Cu(100).

| Fixed Analyzer Energy ($KE_1$) | $KE_{2max}$ |
|---|---|
| 136.25eV (Core) | 67.25eV |
| 150eV | 53.5eV |
| 165eV | 38.5eV |
| 180eV | 23.5eV |
| 190eV | 13.5eV |
| 197eV | 6.5eV |

*Table 1. Fixed analyzer energies with their corresponding $E_S$ values for Cu(100). It can be seen that 67.25eV of electron energy contributes towards the LET region of the Cu 3p-MVV APECS Auger spectrum due to VB-VB coincidence.*

The background estimation from the *VB-VB* spectra taken at the energies between the core peak and the valence band are extrapolated to determine the background contribution in the APECS Auger spectrum. The topmost red curve in Fig. 2-f represents an estimate of *VB-VB* contributions to the APECS spectra based upon an extrapolation of the measured VB-VB spectra shown in Fig. 2(a-e). In Fig. 3, we show the comparison of Auger spectrum before and after the subtraction of the measured background. The normalized count rates significantly drops by more than half, from 45 to less than 20. The more prominent Auger line shape extends from 50eV to 65eV, now entailing a smaller LET that extends to 0eV.

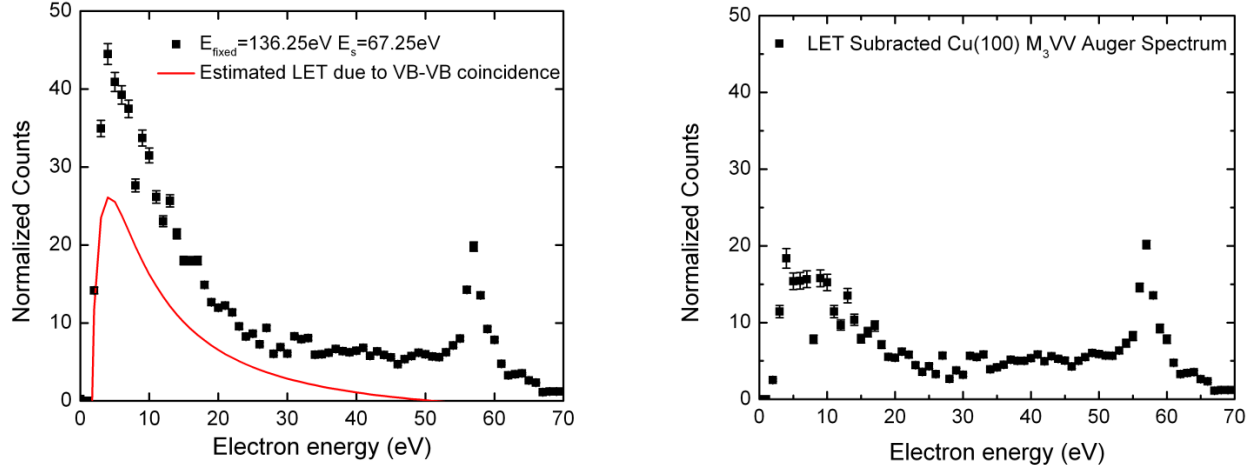

fig. 3 (left panel) *Estimation of the VB-VB coincidence spectra (red) from the measured background spectra and APECS Spectrum for Cu(100). (Right panel) APECS Spectrum after subtracting the measured VB-VB contributions.*

**IV. SUMMARY**

We have developed an experimental method to determine the contribution to the LET region in APECS spectrum from photoemission processes in which the energy of the incident photon is shared by two or more VB electrons (*VB-VB* coincidence processes) and one of the photoelectrons is detected in the same energy range as the selected core electron. We have determined that VB-VB processes can contribute a significant background to the LET of APECS spectra that cannot be removed using coincidence methods alone. We determined both the magnitude and spectral distribution of this background by making the measurements in which one of the electrons was in the energy range below the valence band peak and above the core level photo-emission energy. In this range, the coincidence requirement precludes the detection of electrons emitted as a result of Auger transitions. Specifically, a beam of 200eV photons was incident on a Cu(100) sample and a series of coincidence measurements were performed using a spectrometer equipped with two cylindrical mirror analyzers (CMAs). One CMA was set at series of fixed energies that ranged between the energy of the core and the VB peaks and the other CMA was scanned over a range corresponding to electrons leaving the surface between 0eV and 70eV. A parameterized function was used to fit the spectra obtained from the coincidence measurement of the inelastically scattered VB electrons and fixed energies chosen between the core peak and the VB. Extrapolated parameters based on these fitting curves were used to find the estimation of the secondary electrons contribution to the APECS Auger spectrum.

It is seen that (see Fig. 3) the background due to VB-VB coincidences constitutes only a few percent of the signal in the Auger peak region. It suggests that the background is not of significance in measurements confined to the Auger peak region. However, we found that the VB-VB coincidence background contributes as much as 50% of the APECS spectra in the low energy "secondary peak" region. A final Auger spectrum was obtained by subtracting the Auger unrelated VB-VB processes from the APECS spectrum to obtain the spectrum of electrons that are emitted solely as a result of Auger transition processes initiated by the creation of the $3p_{3/2}$ core hole.

*Acknowledgement*: The authors would like to thank *the department of physics*, University of Texas at Arlington for financial support and the referees of this paper for their comments and suggestions. This research was sponsored in part by the NSF DMR 0907679. Use of the National Synchrotron Light Source, Brookhaven National Laboratory, was supported by the U.S. Department of Energy, Office of Science, Office of Basic Energy Sciences, under Contract No. DE-AC02-98CH10886.